\documentclass[a4paper]{jpconf}

\usepackage{graphicx}
\usepackage{amsmath}

\begin{document}

\title{Improved solution to CMB quadrupole problem using ellipsoidal Universes with Chaplygin gas}

\author{Mohamed Lamine Abdelali and Noureddine Mebarki}

\address{Laboratoire de Physique Mathematique et Subatomique, Universite Freres Mentouri Constantine 1, BP 325 Route de Ain El Bey, Constantine 25017, Algeria}

\ead{mabdelali1@gmail.com}

\begin{abstract}
A Universe containing uniform magnetic fields, strings, or domain walls is shown to have an ellipsoidal expansion. 
This case has motivations from observational cosmology especially the anomaly concerning the low quadrupole amplitude compared to the best-fit $\Lambda$CDM prediction in Planck data. 
It is shown that a Universe with eccentricity at decoupling of order $10^{-2}$ can reduce the quadrupole amplitude without affecting higher multipoles of the angular power spectrum of the temperature anisotropy. 
We study the evolution of ellipsoidal Universes using dynamical system techniques for the first time. 
The determined critical points vary between saddle and past attractors depending on dark energy state equation parameter $w_{\Lambda}$, with no future attractors. 
Important results are shown with numerical integrations of this dynamical system done using several initial conditions. 
For instance, a tendency for high expansion differences between planar and perpendicular axes is observed which contradicts previous assumption on the evolution behaviour of ellipsoid Universes. 
Considering dark energy as a Chaplygin gas solves this contradiction by controlling cosmic shear evolution. 
\end{abstract}

\section{Introduction}

Our understanding of the Universe in modern cosmology is highly influenced by the discovery of Cosmic Microwave Background (CMB). 
The CMB angular power spectrum is measured by a succession of ground-based and especially satellite missions, such as Cosmic Background Explorer (COBE) satellite (see \cite{RId006.1}), Wilkinson Microwave Anisotropy Probe (WMAP) (see \cite{RId006.2}, \cite{RId008.1}, \cite{RId008.2} and \cite{RId002.1}) and recently by the Planck satellite (see \cite{RId008.3}). 
Those observations confirm the standard Big Bang model. 
In this model and at large scales, our Universe is isotropic and homogeneous described by Friedmann-Lemaitre-Robertson-Walker (FLRW) line element. 
Small deviations from isotropy are also reported by these three satellite missions and with agreement to predictions of the $\Lambda$CDM model. 
Since 1992 with the first COBE mission, some anomalous features in CMB anisotropies are reported and confirmed since then repeatedly by the two following missions (see COBE results \cite{RId005.1}, \cite{RId001.7}, WMAP results \cite{RId001.1}, \cite{RId001.2}, \cite{RId001.3}, \cite{RId001.4}, \cite{RId001.5} and Planck results \cite{RId008.3}). 
These peculiarities reside at large scales which are mainly : suppression of power for large angular scales especially quadrupole moment, unusual alignment between axes of quadrupole and octopole moments, non-Gaussian signatures caused by observed cold spot and an hemispherical power asymmetry for large scales. 
The quadrupole problem is the most important discrepancy between the measured value and the best-fit $\Lambda$CDM model. 
For instance for Planck data (see \cite{RId008.3}), the observed quadrupole moment is $299.5\,\mu K^2$ while the expected $1150 \pm 727\,\mu K^2$. 

Since the first evidence of COBE, studies were done to evaluate the significance of these anomalies (see \cite{RId001.1}, \cite{RId001.2}, \cite{RId001.3}, \cite{RId001.4}, \cite{RId001.5} and \cite{RId001.8}). 
Although, the probability of low quadrupole is not statistically significant (see \cite{RId002.1}) since at the largest angular scales, measurements are affected by foreground and other systematic effects. 
The succession of missions did not remove this observed effect. 
These features captured attention especially if this discrepancy has a cosmological origin. 
Measurements of polarization provide possible indications on signatures of cosmological anisotropy (see \cite{RId003.9}, \cite{RId003.10} and \cite{RId003.11}). 
Also, it may be signal of a nontrivial topology (see \cite{RId006.8}, \cite{RId002.2}, \cite{RId002.3} and \cite{RId002.4}). 
In the literature, several proposals are advanced to account for those anomalies (see \cite{RId001.5}, \cite{RId005.8} and \cite{RId002.5}). 
An interesting approach to solve the quadrupole problem is to use simple modification to conventional FLRW model by considering a possible small global anisotropy in the Universe. 
The quadrupole and octopole alignment defines unique directions : one longitudinal and two transverse directions with respect to an equatorial plane. 
Other evidence for the symmetry axis come from analysis of spiral galaxies in the Sloan Digital Sky Survey (see \cite{RId003.6}) and polarization of light travelling over cosmic distances (see \cite{RId003.7} and \cite{RId003.8}). 
A substitute to isotropic and homogeneous FLRW model is a spatially homogeneous but anisotropic cosmological model with plane-symmetric line element known as ellipsoidal Universe (see \cite{RId006.9} and \cite{RId006.10}). 
A good replacement of the standard model of cosmology should reproduce the succession of dominance eras, i.e. inflation, radiation, matter and lastly cosmological term (or dark energy in general). 
The energy-momentum tensor of such Universe is spatially non-spherical or spontaneously transform to non-spherical. 
This asymmetry affects all moments of CMB multipole expansion (see \cite{RId001.13} and \cite{RId005.12}). 

The origin of such planar symmetry could be from several mechanisms (see \cite{RId006.10} and \cite{RId005.12}) : topological defects (such as cosmic stings, cosmic domain walls) (see \cite{RId005.12} and \cite{RId001.10}), a uniform cosmic magnetic field (see \cite{RId006.9}, \cite{RId006.10} and \cite{RId001.9}), magnetic fields having planar symmetry on cosmic scales (see \cite{RId006.12}), or a moving dark energy (see \cite{RId006.13}). 
In particular, magnetic fields seems to be the ideal candidate as a source of this anisotropy as supported by several recent observations proving the presence of these fields on cosmic scales (see  \cite{RId001.9}, \cite{RId005.13}, \cite{RId005.14},and \cite{RId001.11}). 
Cosmic magnetic fields have several proposed genesis mechanisms from early eras of Universe evolution (e.g. inflation \cite{RId005.16} and phase transition \cite{RId001.14}) and may affect then the expansion of the universe (see \cite{RId001.12} and \cite{RId005.15}). 
It is estimated that initial magnetic fields can be within an order of magnitude of the critical density in electroweak phase transition (see \cite{RId001.16}). 
In several recent papers (see \cite{RId006.9}, \cite{RId006.10}, \cite{RId006}, \cite{RId008.7} and \cite{RId008}), It has been proposed that ellipsoidal Universe with a small eccentricity at decoupling of order $e_{dec} \sim 0.67 \times 10^{-2}$ produce in a reduction of expected quadrupole moment with respect to best-fit $\Lambda$CDM model and match the low value observed. 
This result is independent of the anisotropy origin and with the low eccentricity, the higher multipoles are not affected. 
The isotropy observed in present day $t_0$ is reproduced by the condition $e(t_0) = 0$ since the eccentricity evolves according to the decreasing magnetic energy density. 

The use of dynamical systems to study cosmological models has joined to well established research areas in physics and mathematics. 
This practice has proven its important aspects in recent years (see \cite{RId010}). 
The abstraction of any system to a dynamical one with an evolution described by a set of equations within a phase space gives us strong indications of its start and fate. 
Then and to study the properties of ellipsoidal Universes, we propose in this paper a dynamical study of such Universe. 
In section \ref{sec.2}, we start by a short presentation and reproduction of key equations describing the evolution and the construction of the dynamical system. 
In section \ref{sec.3}, critical points are determined and their nature studied. 
In section \ref{sec.4}, a numerical integration is done to follow the evolution of such Universe. 
In section \ref{sec.5}, results form an elliptical Universe with Chaplygin gas as dark energy is presented. 
Finally, we close by discussions and conclusions\ref{sec.6}. 

\section{Dynamical system of an ellipsoidal Universe}
\label{sec.2}

For an ellipsoidal Universe, Taub line element (see \cite{RId001.13}) is used to describe the geometry of its spacetime

\begin{equation}
ds^{2} = dt^{2} - a^{2}(t) (dx^{2} + dy^{2}) - b^{2}(t) dz^{2},
\label{eq.6.1}
\end{equation}

where $a(t)$ and $b(t)$ describe the scale factor along different axis. 
The spatial spherical geometry of a flat FLRW spacetime is reproduced momentarily or permanently if $a(t)=b(t)$. 
The source of the anisotropy of ellipsoidal Universe is related to anisotropy in the pressure of a particular Universe component. 
The energy momentum tensor of such Universe is given by 

\begin{equation}
T^{\mu}_{\nu} = diag(\rho, -p_{\parallel}, -p_{\parallel}, -p_{\perp}),
\end{equation}

where $\rho$ is the energy density of usual isotropic components and anisotropic components $\rho = \rho ^I +\rho ^A$. 
The pressure is different for the planar axis $p_{\parallel}$ and for the orthogonal axis $p_{\perp}$. 
They are also a sum of isotropic and anisotropic components $p_{\parallel} = p^I + p^A_{\parallel}$ and $p_{\perp} = p^I + p^A_{\perp}$. 

The usual isotropic components of the Universe are matter ($\rho _m$, $p_m=0$), radiation ($\rho _r$, $p_r=\frac{1}{3} \rho _r$) and dark energy ($\rho _{\Lambda}$, $p_{\Lambda}=w_{\Lambda}\rho _{\Lambda}$). 
For the anisotropic component $\rho ^A$, the two pressure components have two different state equation parameters ($p^A_{\parallel}=w_1 \rho ^A$) and ($p^A_{\perp}=w_3 \rho ^A$). 
Similarly to FLRW spacetime, we define Hubble parameter and density parameters, but using average scale factor $A = (a^2(t) b(t))^{1/3}$. 
The cosmological parameters are then given by, starting with Hubble parameter 

\begin{equation}
H = \frac{\dot{A}}{A},
\end{equation}

critical density 

\begin{equation}
\rho _c = \frac{3 H^2}{8 \pi G}
\end{equation}

and density parameters

\begin{equation}
\Omega _i = \frac{\rho _i}{\rho _c}.
\end{equation}

where $i$ represent the different Universe components matter $m$, radiation $r$, dark energy $\Lambda$ and the source of anisotropy $A$. 
To describe the expansion rate of the Universe in each axis, we also define 

\begin{equation}
H _a = \frac{\dot{a}}{a},
\end{equation}

and 

\begin{equation}
H _b = \frac{\dot{b}}{b}.
\end{equation}

The anisotropy of this Universe is represented by two quantities, eccentricity $e$ and cosmic shear $\Sigma$ given by 

\begin{equation}
\begin{split}
e = \sqrt{1-(\frac{b}{a})^2} \,\,...\,\,a>b \\
\,\,\, \sqrt{1-(\frac{a}{b})^2} \,\,...\,\,b>a,
\end{split}
\end{equation}

and 

\begin{equation}
\Sigma _a = \Sigma = \frac{H _a - H}{H}.
\label{eq.6.3}
\end{equation}

The first describes difference in scale factors between axis and the second describes the difference in expansion rate between axis. 

Using Taub line element, energy momentum tensor and Einstein equations, we obtain the temporal evolution of both scale factors $a(t)$ and $b(t)$ given by 

\begin{equation}
(\frac{\dot{a}}{a})^{2} +2(\frac{\dot{a}}{a}) (\frac{\dot{b}}{b}) = 8\pi G (\rho ^{I} + \rho ^{A}),
\label{eq.6.4}
\end{equation}

\begin{equation}
(\frac{\ddot{a}}{a}) +(\frac{\ddot{b}}{b}) +(\frac{\dot{a}}{a}) (\frac{\dot{b}}{b}) = - 8\pi G (p ^{I} + p ^{A}_{\parallel})
\label{eq.6.5}
\end{equation}

and 

\begin{equation}
2(\frac{\ddot{a}}{a}) +(\frac{\dot{a}}{a})^{2} = - 8\pi G (p ^{I} + p ^{A}_{\perp})
\label{eq.6.6}
\end{equation}

The energy momentum tensor conservation $T^{\mu}_{\nu ; \mu} = 0$ describes the temporal evolution of densities. 
It is considered (see \cite{RId006.9}) that the isotropic and anisotropic components are conserved separately $(T_I)^{\mu}_{\nu ; \mu} = 0$ and $(T_A)^{\mu}_{\nu ; \mu} = 0$. 
Then, we obtain the following equations 

\begin{equation}
\dot{\rho} _m = - 3 H \rho _m,
\label{eq.6.8}
\end{equation}

\begin{equation}
\dot{\rho} _r = - 4 H \rho _r,
\label{eq.6.9}
\end{equation}

\begin{equation}
\dot{\rho} _{\Lambda} = -3H (1+w_{\Lambda}) \rho _{\Lambda}
\label{eq.6.10}
\end{equation}

and 

\begin{equation}
\dot{\rho} ^A = - [2 (1+w_1) H_a + (1+w_3) H_b] \rho ^A.
\label{eq.6.11}
\end{equation}

To this step, we reproduced results from previous literature for Einstein equations of ellipsoidal Universe (e.g. see \cite{RId006.10}). 
To construct the dynamical system needed to study the ellipsoidal Universe, we start by rewriting Einstein equations in function of average Hubble constant and cosmic shear using the fact that 

\begin{equation}
H_a = (1 + \Sigma) H
\label{eq.6.13}
\end{equation}

and 

\begin{equation}
H_b = (1 -2 \Sigma) H.
\label{eq.6.14}
\end{equation}

Then, the first Einstein equation \ref{eq.6.4} is rewritten to be 

\begin{equation}
((1 + \Sigma) H)^{2} +2(1 + \Sigma) H (1 -2 \Sigma) H = 8\pi G (\rho ^{I} + \rho ^{A}).
\label{eq.6.14.1}
\end{equation}

After simplification, we obtain an equation similar to closure equation of FLRW spacetime 

\begin{equation}
1 = \Omega _m + \Omega _r + \Omega _{\Lambda} + \Omega _{A} + \Sigma ^2.
\label{eq.6.15}
\end{equation}

The second equation \ref{eq.6.5} is rewritten to be 

\begin{equation}
\dot{H_a} +\dot{H_b} +H_a^2 +H_b^2 +H_a H_b = - 8\pi G (p ^{I} + p ^{A}_{\parallel}),
\label{eq.6.19}
\end{equation}

and equation \ref{eq.6.6} is rewritten to be 

\begin{equation}
2\dot{H_a} +3 H_a^2 = - 8\pi G (p ^{I} + p ^{A}_{\perp}).
\label{eq.6.20}
\end{equation}

By executing (2 * equation \ref{eq.6.19} + equation \ref{eq.6.20}), we obtain in function of the average Hubble parameter and cosmic shear 

\begin{equation}
\frac{\dot{H}}{H^2} = - \frac{3}{2} (1+\Sigma ^2) - \frac{1}{2} ( \Omega _r + 3 w_{\Lambda} \Omega _{\Lambda} + (2 w_1 + w_3) \Omega _{A}).
\label{eq.6.21}
\end{equation}

By introducing the temporal evolution of density parameters given by 

\begin{equation}
\dot{\Omega} _i = \frac{\dot{\rho} _i}{\rho _c} - \Omega _i \frac{\dot{\rho} _c}{\rho _c},
\end{equation}

or equivalently by 

\begin{equation}
\dot{\Omega} _i = \frac{\dot{\rho} _i}{\rho _c} - 2 \Omega _i \frac{\dot{H}}{H}.
\end{equation}

We obtain the following equations for each component of the Universe 

\begin{equation}
\dot{\Omega} _m = -3 H \Omega _m + H \Omega _m [3+3\Sigma ^2 + \Omega _r + 3 w_{\Lambda} \Omega _{\Lambda} + (2 w_1 + w_3) \Omega _{A}],
\label{eq.6.23}
\end{equation}

\begin{equation}
\dot{\Omega} _r = -4 H \Omega _r + H \Omega _r [3+3\Sigma ^2 + \Omega _r + 3 w_{\Lambda} \Omega _{\Lambda} + (2 w_1 + w_3) \Omega _{A}],
\label{eq.6.24}
\end{equation}

\begin{equation}
\dot{\Omega} _{\Lambda} = -3H (1+w_{\Lambda}) \Omega _{\Lambda} + H \Omega _{\Lambda} [3 + 3\Sigma ^2 + \Omega _r + 3 w_{\Lambda} \Omega _{\Lambda} + (2 w_1 + w_3) \Omega _{A}]
\label{eq.6.25}
\end{equation}

and

\begin{equation}
\dot{\Omega} _A = - [2 (1+w_1) (1+\Sigma) + (1+w_3) (1-2\Sigma)] H \Omega _A + H \Omega _A [3+3\Sigma ^2 + \Omega _r + 3 w_{\Lambda} \Omega _{\Lambda} + (2 w_1 + w_3) \Omega _{A}].
\label{eq.6.26}
\end{equation}

As expressed in equation \ref{eq.6.15}, we need to obtain evolution equation of cosmic shear. 
From equations \ref{eq.6.3} and \ref{eq.6.20}, we obtain 

\begin{equation}
\dot{\Sigma} = - 3H [\frac{1}{2} (1+\Sigma)^2 +\frac{\Omega _r}{3} +w_{\Lambda} \Omega _{\Lambda} +w_3 \Omega _A]
 +H (1+\Sigma) [ \frac{3}{2} (1+\Sigma ^2) + \frac{1}{2} ( \Omega _r + 3 w_{\Lambda} \Omega _{\Lambda} + (2 w_1 + w_3) \Omega ^{A})].
\label{eq.6.27}
\end{equation}

By using the fact that $\dot{\Omega} = H \Omega '$, the dynamical system of equations describing the ellipsoidal Universe is given by 

\begin{equation}
\Omega _m ' = 3 \Omega _m [\Sigma ^2 + \frac{1}{3} \Omega _r +w_{\Lambda} \Omega _{\Lambda} + \frac{2 w_1 + w_3}{3} \Omega _{A}],
\label{eq.6.28}
\end{equation}

\begin{equation}
\Omega _r ' = 3 \Omega _r [-\frac{1}{3} + \Sigma ^2 + \frac{1}{3} \Omega _r +w_{\Lambda} \Omega _{\Lambda} + \frac{2 w_1 + w_3}{3} \Omega _{A}],
\label{eq.6.29}
\end{equation}

\begin{equation}
\Omega _{\Lambda} ' = 3 \Omega _{\Lambda} [ - w_{\Lambda}) + \Sigma ^2 + \frac{1}{3} \Omega _r +w_{\Lambda} \Omega _{\Lambda} + \frac{2 w_1 + w_3}{3} \Omega _{A}],
\label{eq..6.30}
\end{equation}

\begin{equation}
\Omega _A ' = 3 \Omega _A [-\frac{[2 (1+w_1) (1+\Sigma) + (1+w_3) (1-2\Sigma)]}{3} + 1 + \Sigma ^2 + \frac{1}{3} \Omega _r +w_{\Lambda} \Omega _{\Lambda} + \frac{2 w_1 + w_3}{3} \Omega _{A}]
\label{eq.6.31}
\end{equation}

and

\begin{equation}
\Sigma ' = - 3 [\frac{1}{2} (1+\Sigma)^2 +\frac{\Omega _r}{3} +w_{\Lambda} \Omega _{\Lambda} +w_3 \Omega _A
 + \frac{1}{2} (1+\Sigma) (1+\Sigma ^2 + \frac{1}{3} \Omega _r + w_{\Lambda} \Omega _{\Lambda} + \frac{(2 w_1 + w_3)}{3} \Omega _{A})].
\label{eq.6.32}
\end{equation}

To verify the closure equation \ref{eq.6.15}, we impose the following condition on density parameters and cosmic shear 

\begin{equation}
0 \leq \Omega _i \leq 1
\label{eq.6.32.1}
\end{equation}

and 

\begin{equation}
-1 \leq \Sigma \leq 1
\label{eq.6.32.2}
\end{equation}

From the same equation, we could reduce this set of equations to a system of only 4 degrees of freedom when writing $\Omega _m$ in function of the rest of cosmological parameters. 

\section{Nature of critical points}
\label{sec.3}

The non-linearity of the dynamical system obtained in the previous section impose the use of symbolic solvers available in Python and exactly in Sympy.Nonlinsolve. 
This solver is used for this set of equations for each anisotropy source: magnetic fields ($w_1 = 1$, $w_3 = -1$), walls ($w_1 = -1$, $w_3 = 0$) and strings ($w_1 = 0$, $w_3 = -1$). 
The critical points found are in function of $w _\Lambda$ of dark energy. 
Not all critical points given by the solver are accepted as they should verify the physical constrains in equations \ref{eq.6.32.1} and \ref{eq.6.32.2}. 
For solutions depending on $w_{\Lambda}$, we solve the inequality for the corresponding cosmological parameter to find the range of accepted $w_{\Lambda}$. 

For the case of magnetic fields, the critical points are 

\begin{enumerate}
\item[M.1.] (-72, 0, 0, 5),
\item[M.2.] ($\frac{-9}{32}$, 0, $\frac{81}{32}$, $\frac{1}{2}$), 
\item[M.3.] (0, 0, 0, -1), 
\item[M.4.] (0, 0, 0, $\frac{-1 -I \sqrt{7}}{2}$), 
\item[M.5.] (0, 0, 0, $\frac{-1 +I \sqrt{7}}{2}$), 
\item[M.6.] (0, 0, $\frac{42}{25}$, $\frac{1}{5}$), 
\item[M.7.] (0, $-3\frac{(5w_{\Lambda} - 3)^2}{(128 w_{\Lambda})}$, $3\frac{(3 w_{\Lambda}^2 + 86 w_{\Lambda} + 19)}{128}$, $\frac{3 w_{\Lambda} - 1}{4}$), 
\item[M.8.] (0, $\frac{(-w_{\Lambda}^2 - 10w_{\Lambda} - 13 + (w_{\Lambda} + 3 ) \sqrt{w_{\Lambda}^2 + 18 w_{\Lambda} + 17} )}{(2 w_{\Lambda})}$, 0, $\frac{w_{\Lambda} + 3 - \sqrt{w_{\Lambda}^2 + 18 w_{\Lambda} + 17} + 3}{2}$), 
\item[M.9.] (0, $-\frac{(w_{\Lambda}^2 + 10 w_{\Lambda} + 13 + (w_{\Lambda}+3) \sqrt{w_{\Lambda}^2 + 18 w_{\Lambda} + 17} )}{(2*w_{\Lambda})}$, 0, $\frac{w_{\Lambda} + 3 + \sqrt{w_{\Lambda}^2 + 18 w_{\Lambda} + 17}}{2}$).
\end{enumerate}

For this case, only the critical point (M.3) verify the physical constrains and with no condition on $w_{\Lambda}$. 
The eigenvalues of this critical point are $(1, 2, 2/3, 1-w_{\Lambda})$ representing a past attractor or a saddle point depending on the value of $w_{\Lambda}$. 

For the case of walls, the critical points are 

\begin{enumerate}
\item[W.1.] (-72, 0, 0, 5),
\item[W.2.] (0, 0, $\frac{3}{2}$, -1), 
\item[W.3.] (0, 0, 0, -1), 
\item[W.4.] (0, 0, 0, $\frac{-1 -I \sqrt{7}}{2}$), 
\item[W.5.] (0, 0, 0, $\frac{-1 +I \sqrt{7}}{2}$), 
\item[W.6.] (0, 0, 21, -4), 
\item[W.7.] ($\frac{9}{8}$, 0,  $\frac{135}{16}$, $\frac{-5}{2}$), 
\item[W.8.] (0, $\frac{(-w_{\Lambda}^2 - 10 w_{\Lambda} - 13 + (w_{\Lambda} +3) \sqrt{w_{\Lambda}^2 + 18 w_{\Lambda} + 17})}{(2 w_{\Lambda})}$, 0, $\frac{w_{\Lambda} + 3 - \sqrt{w_{\Lambda}^2 + 18 w_{\Lambda} + 17}}{2}$), 
\item[W.9.] (0, $-\frac{(w_{\Lambda}^2 + 10 w_{\Lambda} + 13 + (w_{\Lambda} +3) \sqrt{w_{\Lambda}^2 + 18 w_{\Lambda} + 17})}{(2 w_{\Lambda})}$, 0, $\frac{w_{\Lambda} + 3 + \sqrt{w_{\Lambda}^2 + 18 w_{\Lambda} + 17}}{2}$), 
\item[W.10.] (0, $-3 \frac{(w_{\Lambda} - 2)}{8}$, $3\frac{(3(w_{\Lambda} - 2) + 2)(5(w_{\Lambda} - 2) + 4)}{16}$, $-3\frac{w_{\Lambda}}{2} - 1$).
\end{enumerate}

For this case, the accepted critical points are (W.3) with no condition on $w_{\Lambda}$ and (W.10) for $w_{\Lambda} \in [-2/3, -0.4/3]$. 
The eigenvalues of the critical point (W.3) are $(1, 1, 2/3, 1-w_{\Lambda})$ representing a past attractor or a saddle point depending on the value of $w_{\Lambda}$. 
The eigenvalues of the critical point (W.10) for $w_{\Lambda}=-2/3$ are $(-1, -2/3, 0.00, 7/6)$ representing a saddle point. 
Changing the value of $w_{\Lambda}$ for (W.10) changes the nature of the critical point between a past attractor or a saddle point.

For the case of strings, the critical points are 

\begin{enumerate}
\item[S.1.] (-72, 0, 0, 5),
\item[S.2.] (0, 0, 0, -1), 
\item[S.3.] (0, 0, 0, $\frac{-1 -I \sqrt{7}}{2}$), 
\item[S.4.] (0, 0, 0, $\frac{-1 +I \sqrt{7}}{2}$), 
\item[S.5.] ($\frac{-9}{4}$, 0,  $\frac{27}{4}$, 2), 
\item[S.6.] (0, 0, $\frac{-240 \sqrt{3} + 570}{169}$, $\frac{-6 \sqrt{3} + 11}{13}$), 
\item[S.7.] (0, 0, $\frac{240 \sqrt{3} + 570}{169}$, $\frac{6 \sqrt{3} + 11}{13}$), 
\item[S.8.] (0, $\frac{-3 (13 w_{\Lambda}^2 - 6 w_{\Lambda} - 3)}{(16 w_{\Lambda})}$, $\frac{-3 (3 w_{\Lambda}^2 - 26 w_{\Lambda} - 13)}{16}$, $ \frac{3 w_{\Lambda} + 1}{2}$), 
\item[S.9.] (0, $\frac{(-w_{\Lambda}^2 - 10 w_{\Lambda} - 13 + (w_{\Lambda} +3) \sqrt{w_{\Lambda}^2 + 18 w_{\Lambda} + 17})}{(2 w_{\Lambda})}$, 0, $\frac{w_{\Lambda} + 3 - \sqrt{w_{\Lambda}^2 + 18 w_{\Lambda} + 17}}{2}$), 
\item[S.10.] (0, $-\frac{(w_{\Lambda}^2 + 10 w_{\Lambda} + 13 + (w_{\Lambda} +3) \sqrt{w_{\Lambda}^2 + 18 w_{\Lambda} + 17})}{(2 w_{\Lambda})}$, 0, $\frac{w_{\Lambda} + 3 + \sqrt{w_{\Lambda}^2 + 18 w_{\Lambda} + 17}}{2}$). 
\end{enumerate}

For this case, the accepted critical points are (S.2) and (S.6) with no condition on $w_{\Lambda}$. 
The eigenvalues of the critical point (S.2) are $(1, 2, 2/3, 1-w_{\Lambda})$ representing a past attractor or a saddle point depending on the value of $w_{\Lambda}$. 
The eigenvalues of the critical point (S.6) for $w_{\Lambda}$ are $(0.46-w_{\Lambda}, 0.12, 2.09, -0.18)$ representing a past attractor or a saddle point depending on the value of $w_{\Lambda}$. 
Then, most critical points are refused in the 3 cases of magnetic fields, walls or strings. 
No future attractors are found for the 3 cases. 

\section{Numerical integration of the dynamical system}
\label{sec.4}

The study of the critical points and their nature in the previous section does not give a clear indication of the evolution of ellipsoidal Universe especially with the absence of future attractors. 
A numerical integration would be a suitable approach to study the evolution of such complex and non-linear dynamical system of equations. 
For this purpose, we use forward Euler method where the derivative of each parameter $X$ is rewritten to be 

\begin{equation}
X ' = \frac{X_{n+1} - X_{n}}{\Delta t} = f(X_{n}).
\end{equation}

Then, the next step value $X_{n+1}$ is obtained from the past value $X_{n}$ by 

\begin{equation}
X_{n+1} = X_{n} + f(X_{n}) \times \Delta t,
\end{equation}

\begin{figure}[h]
\includegraphics[width=40pc]{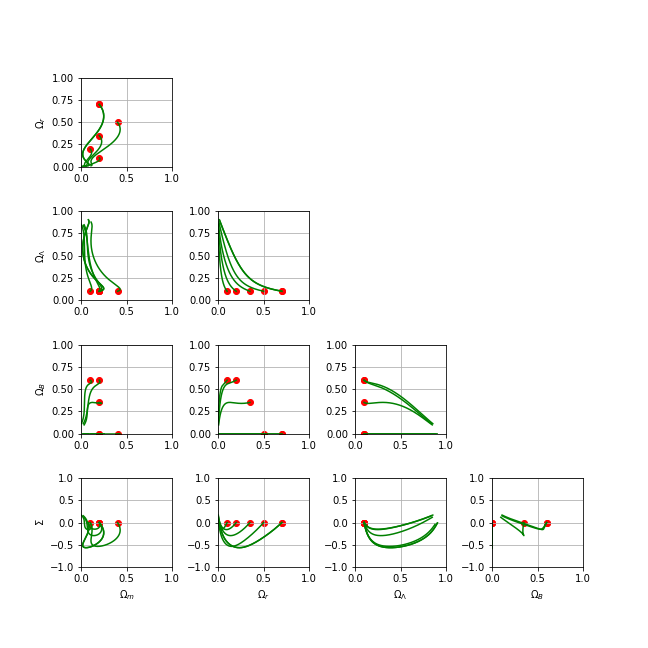}\hspace{2pc}%
\caption{\label{fig.1}Cross plotting of cosmological parameters and comic shear evolution using numerical integration : cases with null initial $\Sigma$.}
\end{figure}

\begin{figure}[h]
\includegraphics[width=40pc]{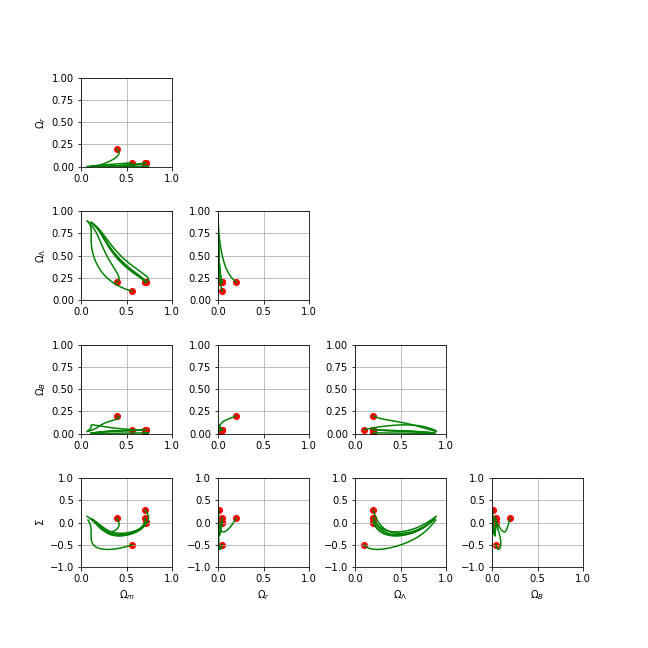}\hspace{2pc}%
\caption{\label{fig.2}Cross plotting of cosmological parameters and comic shear evolution using numerical integration : cases with different initial $\Sigma$.}
\end{figure}

\begin{figure}[h]
\includegraphics[width=40pc]{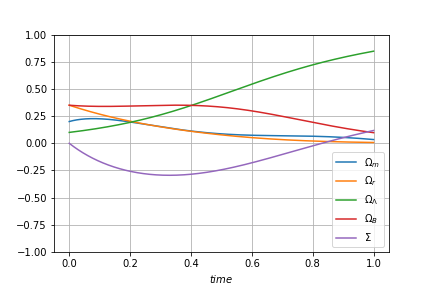}\hspace{2pc}%
\caption{\label{fig.3}Temporal evolution of cosmological parameters and comic shear evolution using numerical integration.}
\end{figure}

When applying this principle on all parameters present in the dynamical system, we need to start from a single initial combination of ($\Omega _r$, $\Omega _{\Lambda}$, $\Omega _A$, $\Sigma$). 
Then, this numerical method follows the evolution step by step. 
We fixed the integration step $\Delta t$ to the smallest value possible of $10^{-2}$ to maximize accuracy without affecting computational efficiency. 
The numerical integration is interrupted if one parameter violate the physical conditions. 

To represent results of this method, we use several plots given in figure \ref{fig.1} and \ref{fig.2}. 
In those figures, the red dots represent the initial position of the Universe and the green trajectory represents the Universe evolution. 
Those figure contain several plots illustrating the evolution and cross dependence of each parameter to all other parameters. 
Those represented combinations are for ellipsoidal universes with magnetic fields as the source of anisotropy. 
But, the results of this case are reproduced for the two other cases (Walls and Strings). 
We represent several initial combinations to observe the general pattern of evolution. 
In those combinations, we vary the universe composition ($\Omega _r$, $\Omega _{\Lambda}$, $\Omega _A$) (see figure \ref{fig.1}) with null cosmic shear. 
We vary also the cosmic shear value $\Sigma$ (see figure \ref{fig.2}). 

Both matter density and radiation density parameters tend toward lower contributions with the universe evolution. 
Inversely, dark energy density parameter evolution is toward higher contribution. 
This repeated pattern of evolution is a reproduction of FLRW evolution in which the dominance is for radiation, then matter and finally dark energy. 
This gives an argument in favour of ellipsoidal Universe to be a good approach to FLRW evolution patterns (see also Figure \ref{fig.3}). 
For the anisotropic component (magnetic fields), the density parameter tend to be lower with the evolution of the Universe. 

The most important observation is for the cosmic shear describing the anisotropy in cosmic expansion. 
Cosmic shear tend to reach -0.5 representing a large difference in cosmic expansion in the planar axis and the orthogonal axis up to ($H_b =4H_a$). 
This is reproduced for different initial values and even with null initial values as shown in all figures. 
This large difference in cosmic expansion between axis and for long periods of the Universe evolution creates a highly anisotropic Universe. 
This result contradicts the key assumption that a small initial eccentricity of the Universe of order $10^{-2}$ would tend to vanishing reproducing the spatial spherical geometry of FLRW Universe. 
This is affecting the use of ellipsoidal Universes as solutions to the power suppression in CMB. 

\section{Case of Generalized Chaplygin Gas}
\label{sec.5}

To overcome the unwanted large cosmic shear, we propose to change the perspective of composition of the Universe. 
The main components used in the model of the previous section are : matter, radiation, dark energy as a cosmological constant and anisotropic component such as magnetic fields. 
The only component which could be reconsidered is dark energy nature since matter and radiation are proven by experiments and magnetic fields are needed to create the Universe anisotropy. 
To change the nature of dark energy, a much general form of state equation should used. 
This state equation is given by 
\begin{equation}
\dot{\rho} _{DE} = -3H (1+\frac{P _{DE}}{\rho _{DE}} ) \rho _{DE}.
\end{equation}
This change would affect all evolution equations of cosmic components since all equations are coupled. 
Except the closure equation which remain the same given by 
\begin{equation}
1 = \Omega _m + \Omega _r + \Omega _{DE} + \Omega _{B} + \Sigma ^2.
\end{equation}

Because of this general form of dark energy, the phase space of the dynamical system describing the evolution of this Universe has another degree of freedom which is critical density. 
It should include another equation for the evolution of critical density. 
The new set of equations is then
\begin{equation}
\Omega _m ' = \Omega _m [3 \Sigma ^2 +  \Omega _r + (2 w_1 + w_3) \Omega _{B} +3 \frac{P _{DE}}{\rho _{Cr}}],
\end{equation}

\begin{equation}
\Omega _r ' = \Omega _r [-1 + 3\Sigma ^2 + \Omega _r + (2 w_1 + w_3) \Omega _{B} +3 \frac{P _{DE}}{\rho _{Cr}}],
\end{equation}

\begin{equation}
\Omega _{DE} ' = \Omega _{DE} [ 3\Sigma ^2 + \Omega _r + (2 w_1 + w_3) \Omega _{B} +3 (\frac{P _{DE}}{\rho _{Cr}} - \frac{P _{DE}}{\rho _{DE}})],
\end{equation}

\begin{equation}
\Omega _B ' = \Omega _B [-(2 (1+w_1) (1+\Sigma) + (1+w_3) (1-2\Sigma)) + 3 + 3\Sigma ^2 + \Omega _r + (2 w_1 + w_3) \Omega _{B} +3 \frac{P _{DE}}{\rho _{Cr}} ],
\end{equation}

\begin{equation}
\Sigma ' = - 3 [\frac{1}{2} (1+\Sigma)^2 +\frac{\Omega _r}{3} +w_3 \Omega _B + \frac{P _{DE}}{\rho _{Cr}}]
 + \frac{(1+\Sigma)}{2} [3+3\Sigma ^2 + \Omega _r  + (2 w_1 + w_3) \Omega _{B} +3 \frac{P _{DE}}{\rho _{Cr}}]
\end{equation}

and

\begin{equation}
\rho _{Cr} ' = -\rho _{Cr} [3 +3 \Sigma ^2 +  \Omega _r + (2 w_1 + w_3) \Omega _{B} +3 \frac{P _{DE}}{\rho _{Cr}}].
\end{equation}

This is complicating further the system in being non linear. 
Then, even symbolic solvers used in previous sections can not find critical points and study their nature. 

To be able to study further this case, one should consider a specific model of dark energy with a specified relation between dark energy pressure and density. 
There are a variety of models of dark energy to replace the cosmological constant in literature. 
Chaplygin gas was introduced in this context and as an alternative to quintessence (e.g. see \cite{RId012.22}). 
Another proposition was to consider a unification between dark matter and dark energy using generalized Chaplychin gas (gCg) (e.g. see \cite{RId012.23}). 
Modified Chaplygin gas (mCg) was also introduced in the framework of FLRW (e.g. see \cite{RId012.24}). 
Since then, several versions of Chaplygin gas are proposed to account further to cosmological observations. 
See \cite{RId012} for recent list of those models. 
A discussion of successes and problems of the introduction of Chaplygin gas can be found in \cite{RId028}. 

Observational constraints on different Chaplygin gas models are obtained using supernovae type Ia, CMB, baryonic acoustic oscillations and Hubble telescope data. 
For gCg, limits on model parameters can be found for example in \cite{RId014}, \cite{RId015} and \cite{RId016}. 
For mCg, constraints can be found in \cite{RId018} and \cite{RId019}. 
The stability of such models are studied with FLRW spacetime (e.g. see \cite{RId020} and \cite{RId021} for gCg, see \cite{RId023} for mCg and see \cite{RId024} for extented Chaplygin gas). 
It is shown that cosmological models with Chaplygin gas does not affect the evolution pattern from radiation dominated to matter dominated to dark energy dominated (see \cite{RId025}). 
Interestingly, some models use gCg as a replacement to cosmological constant without unification with dark matter. 
In such model, the Universe is composed of gCg, CDM, baryons and radiation (see \cite{RId026}). 
Moreover, it is shown that comic walls can emerge in Universe with Chaplygin gas (see \cite{RId027}). 
Cosmic walls can induce an elliptical geometry to the Universe. 
For instance, Chaplygin gas has promising applications and may solve the high cosmic shear values (e.g. see \cite{RId011}). 
Then, it is interesting to study the evolution of such combination: elliptical Universe with Chaplygin gas. 
Instead of dark energy as cosmological constant, we introduce dark energy as a generalized Chaplygin gas. 
The relation between pressure and density in such case is given by 

\begin{equation}
P _{DE} = -\frac{A}{{\rho _{DE}}^{\alpha}}.
\end{equation}

Then, the parts need to be replaced in our set of dynamical equations describing the Universe evolution are 

\begin{equation}
\frac{P _{DE}}{\rho _{DE}} = -\frac{A}{{\rho _{DE}}^{\alpha +1}} = -\frac{A}{{\Omega _{DE}}^{\alpha +1}{\rho _{Cr}}^{\alpha +1}}
\end{equation}

and 

\begin{equation}
\frac{P _{DE}}{\rho _{Cr}} = -\frac{A}{{\rho _{DE}}^{\alpha}\rho _{Cr}} = -\frac{A}{{\Omega _{DE}}^{\alpha}{\rho _{Cr}}^{\alpha +1}}.
\end{equation}

\begin{figure}[h]
\begin{center}
    \includegraphics[width=30pc, height=20pc]{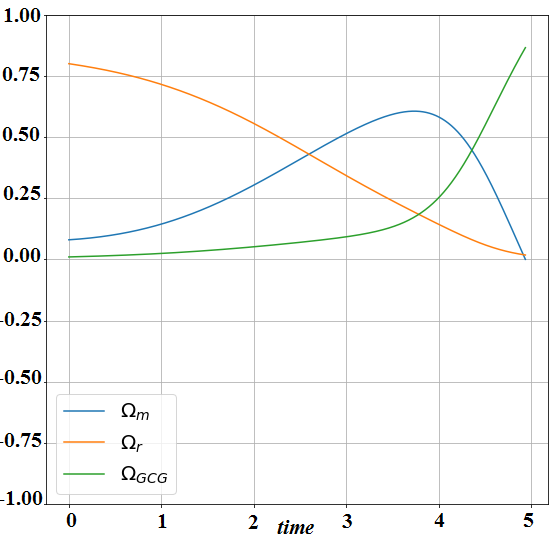}
\end{center}
\caption{\label{fig.4}Temporal evolution of matter, radiation and dark energy density parameters using numerical integration. 
The initial cosmological parameters are $\Omega _m =0.08$, $\Omega _r =0.80$, $\Omega _{DE} =0.01$, $\Omega _B =0.10$, $\Sigma = -0.10$, $A =1.00$, $\alpha =0.02$ and $\rho _{Cr} = 10^7$.}
\end{figure}

\begin{figure}[h]
\begin{center}
    \includegraphics[width=30pc, height=20pc]{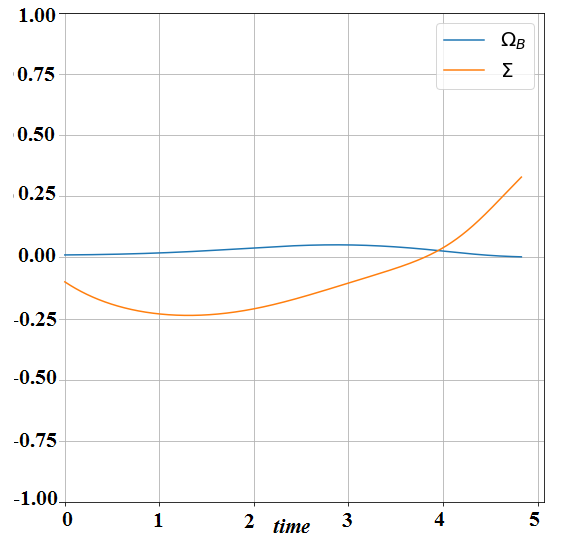}
\end{center}
\caption{\label{fig.5}Temporal evolution of magnetic fields density parameter and comic shear using numerical integration. 
The initial cosmological parameters are $\Omega _m =0.17$, $\Omega _r =0.80$, $\Omega _{DE} =0.01$, $\Omega _B =0.01$, $\Sigma = -0.10$, $A =1.00$, $\alpha =0.02$ and $\rho _{Cr} = 10^7$.}
\end{figure}

\begin{figure}[h]
\begin{center}
    \includegraphics[width=30pc, height=20pc]{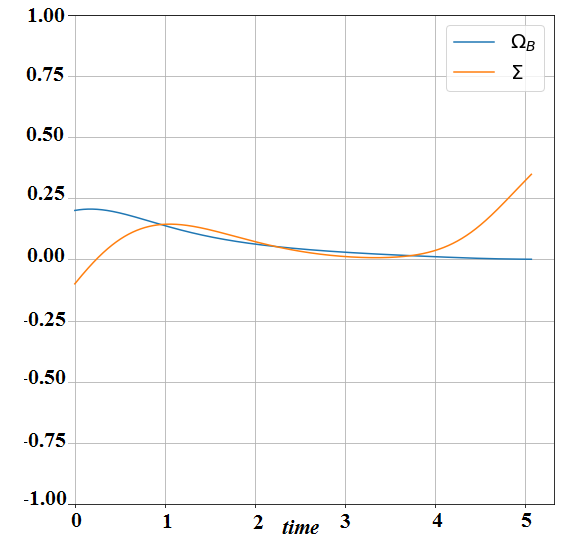}
\end{center}
\caption{\label{fig.6}Temporal evolution of magnetic fields density parameter and comic shear using numerical integration. 
The initial cosmological parameters are $\Omega _m =0.08$, $\Omega _r =0.70$, $\Omega _{DE} =0.01$, $\Omega _B =0.20$, $\Sigma = -0.10$, $A =1.00$, $\alpha =0.02$ and $\rho _{Cr} = 10^7$.}
\end{figure}

\begin{figure}[h]
\begin{center}
    \includegraphics[width=30pc, height=20pc]{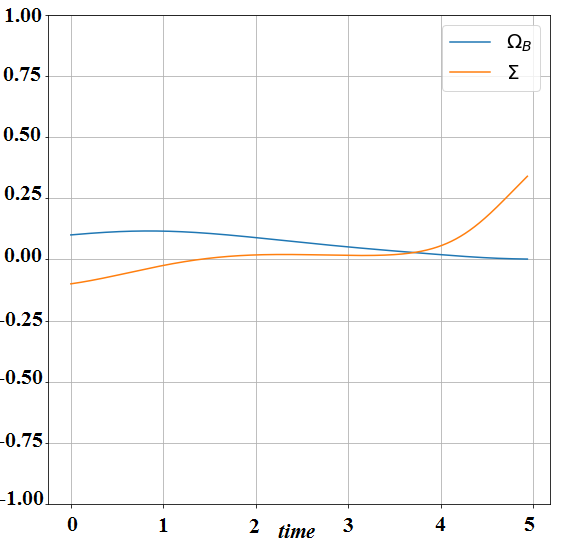}
\end{center}
\caption{\label{fig.7}Temporal evolution of magnetic fields density parameter and comic shear using numerical integration. 
The initial cosmological parameters are $\Omega _m =0.08$, $\Omega _r =0.80$, $\Omega _{DE} =0.01$, $\Omega _B =0.10$, $\Sigma = -0.10$, $A =1.00$, $\alpha =0.02$ and $\rho _{Cr} = 10^7$.}
\end{figure}

The non linearity of this set of equations impose the use of the numerical integration described in the previous section. 
To start the integration, we set the initial values of all density parameters of cosmic components ($\Omega _m$, $\Omega _r$, $\Omega_ {DE}$ and $\Omega _B$), cosmic shear ($\Sigma$) and also of the generalized Chaplygin gas  parameters ($A$ and $\alpha$). 
We use values of gCg parameters favoured by observational constraints ($A=1.0$ and $\alpha 0.02$). 
Following the integration, the pattern of evolution is highly dependent of initial values of those parameters. 
The aim is to find a set of initial values which could reproduce the succession of main Universe components dominance, i.e., radiation then matter then dark energy. 
Such an evolution pattern can be reproduced as shown in figure \ref{fig.4}. 
The second characteristic of a favourable initial combination is a low to zero cosmic shear during evolution to reproduce the observed isotropic expansion of the Universe. 
The simulation of several initial cases has shown a large dependency of the evolution of $\Sigma$ to the initial magnetic fields contribution as it is shown in figures \ref{fig.5}, \ref{fig.6} and \ref{fig.7}. 
Those figures represent 3 Universes with almost the same initial composition of matter, radiation and dark energy. 
The generalized Chaplygin gas also has the same parameters. 
The main change is the magnetic composition. 
A low ($\Omega _B =0.01$) or high ($\Omega _B =0.20$) magnetic fields density parameter has induced a high $\Sigma$ values during long eras of evolution leading then to a high anisotropy Universe. 
Current observations ruled out such cases. 
But with a fine tuned initial magnetic contribution ($\Omega _B = 0.10$) as in figure \ref{fig.7}, the cosmic shear is almost null in most of the Universe evolution. 
In this last case, the expansion is then isotropic in all directions as seen in observations. 

Then, we can conclude that it is possible to find a Universe following the wanted evolution pattern of composition dominance and has an isotropic expansion. 
Such a Universe would have a small initial eccentricity solving the quadrupole problem. 
Also, the cosmic shear is null most of the Universe evolution leading to isotropic expansion. 
This Universe is possible with a particular composition of matter, radiation and dark energy being a generalized Chaplygin gas and having a fine tuned initial magnetic composition and cosmic shear. 
Considering dark energy as a Chaplygin gas solved the large cosmic shear while keeping the explanation of the low CMB quadrupole. 

\section{Discussions and Conclusions}
\label{sec.6}

The $\Lambda$CDM model has several successes in describing our Universe and different observations especially made with more and more precision. 
But, some anomalous measurements still need explanation. 
The quadrupole energy suppression is one of these anomalous observations and is explained as an indication of an initial eccentricity of order $10^{-2}$ at early eras of our Universe. 
Such eccentricity is related with ellipsoidal geometry of the Universe different from FLRW spacetime associated with the $\Lambda$CDM model. 
Authors of this proposed solution to CMB anomaly claim that this eccentricity would vanish during the Universe evolution and then do not affect higher multipoles. 

A dynamical study is performed to verify the properties of ellipsoidal Universe. 
Starting from Taub line element, we reproduced Einstein equations describing the evolution of ellipsoidal Universe. 
To obtain the system of dynamical equations describing this Universe, we has rewriting those equations in function of defined average cosmological parameters. 
We obtain a system of 4 degrees of freedom with a closure equation similar to FLRW model. 
The anisotropy of this Universe is described by the density of the anisotropic component : magnetic fields, walls or strings, contributing by the density parameter of this component. 
The cosmic shear indicates the difference in expansion rate in different Universe axis and also contributes to the dynamical system. 
For each case of anisotropy source, we started by identifying the critical points using symbolic solvers of Python as the system is non-linear. 
The study of critical points gives past attractors and saddle points with no future attractors. 
The nature of those points depends on the value of the dark energy state equation parameter $w_{\Lambda}$.

Another method is used to obtain clear conclusion on the evolution of such ellipsoidal Universe. 
Numerical integration shows that ellipsoidal Universes could reproduce the pattern of radiation, matter then dark energy dominance observed in FLRW spacetimes. 
But, the most important result is about cosmic shear which grows up to $-0.5$. 
Such high values indicate that even small initial eccentricity would evolve to highly anisotropic Universe contrarily to the previous assumption. 
This large anisotropy would affect CMB higher multipoles and diverge the expected values from observed ones which is against the explanation of CMB anomaly with ellipsoidal Universe. 
A change to the model must be made to keep benefits of ellipsoidal Universe in explaining the energy suppression but remove the large cosmic shear. 
Such change is made using dark energy in the form of generalized Chaplygin gas. 
The Universe is then composed of gCg, CDM, baryons and radiations. 
The numerical integration of the dynamical system of such model show a large dependence of cosmic shear evolution to initial magnetic fields components of the Universe. 
Then, a low to null cosmic shear ensuring an isotropic expansion of the Universe as wanted is achievable with a fine tuned initial contribution of magnetic fields. 
The introduction of generalized Chaplygin gas as dark energy solved the unwanted large cosmic shear evolution.
Then, Ellipsoidal universes with generalized Chaplygin gas as dark energy solves the low quadrupole moment of CMB.
Then, further examination of this proposed solution is needed.
The next step would be to constraint the parameters of this model with observational data. 
This possibility is under study and will be subject of future papers. 

\section*{Acknowledgments}

Authors are every indebted to the Algerian of Education and Research (DGRSDT, ATRST) for the financial support.

\end{document}